\documentclass{jpsj-suppl}
\usepackage[utf8]{inputenc}  %Ümläute
\usepackage[T1]{fontenc}
\title{Progress of the Felsenkeller shallow-underground accelerator for nuclear astrophysics}

\author{
D. Bemmerer$^1$,
F. Cavanna$^1$, T.E. Cowan$^{1,2}$, 
M. Grieger $^{1,2}$, T. Hensel $^{1,2}$, A.R. Junghans$^1$, F. Ludwig $^{1,2}$, S. E. Müller$^1$, B. Rimarzig$^1$, S. Reinicke$^{1,2}$,  S. Schulz$^{1,2}$,  R. Schwengner$^1$, K. Stöckel$^2$, T. Szücs$^{1,3}$, M.P. Tak\'acs$^{1,2}$,  A. Wagner$^1$, L. Wagner$^{1,2}$, and  K. Zuber$^2$
}

\inst{$^1$ Helmholtz-Zentrum Dresden-Rossendorf (HZDR), 01328 Dresden, Germany \\
$^2$Technische Universität Dresden (TU Dresden), 01062 Dresden, Germany \\
$^3$MTA ATOMKI, 4026 Debrecen, Hungary
          }
\email{d.bemmerer@hzdr.de}

\recdate{\today}

\abst{%
Low-background experiments with stable ion beams are an important tool for putting the model of stellar hydrogen, helium, and carbon burning on a solid experimental foundation. The pioneering work in this regard has been done by the LUNA collaboration at Gran Sasso, using a 0.4 MV accelerator. In the present contribution, the status of the project for a higher-energy underground accelerator is reviewed. Two tunnels of the Felsenkeller underground site in Dresden, Germany, are currently being refurbished for the installation of a 5\,MV high-current Pelletron accelerator. Construction work is on schedule and expected to complete in August 2017. The accelerator will provide intense, 50\,$\mu$A, beams of $^1$H$^+$, $^4$He$^+$, and $^{12}$C$^+$ ions, enabling research on astrophysically relevant nuclear reactions with unprecedented sensitivity.
}
\begin{document}
\maketitle
%
% ===================================================

\section{Introduction}

Nuclear astrophysics has benefited enormously from data obtained in recent years at the world's only underground ion accelerator, LUNA (Laboratory for Underground Astrophysics) \cite{Broggini10-ARNPS}. This 0.4\,MV accelerator has addressed key reactions of solar fusion, improving the standard solar model  \cite{Adelberger11-RMP}. In addition, several reactions of Big Bang nucleosynthesis \cite{Bemmerer06-PRL,Anders14-PRL} and stellar hydrogen burning \cite{Scott12-PRL,Cavanna15-PRL} have been addressed at LUNA. The reason for this success story is the very low no-beam background. A rock overburden of 1400\,m thickness leads to ultra-low background in $\gamma$-ray detectors \cite{Bemmerer05-EPJA,Caciolli09-EPJA,Szucs10-EPJA}. 

However, the beam energy range is limited, preventing a study of several processes that require higher beam energies, namely the nuclear reactions of helium and carbon burning and the neutron sources of the astrophysical s-process. As a result, there is a call for new, higher-energy underground accelerators \cite{NuPECC10-LRP,WhitePaper2016}. The present contribution reports on the status of the underground accelerator in tunnels VIII and IX of the Felsenkeller site in Dresden, Germany.

% ===================================================
\begin{figure*}[]
\includegraphics[width=\textwidth]{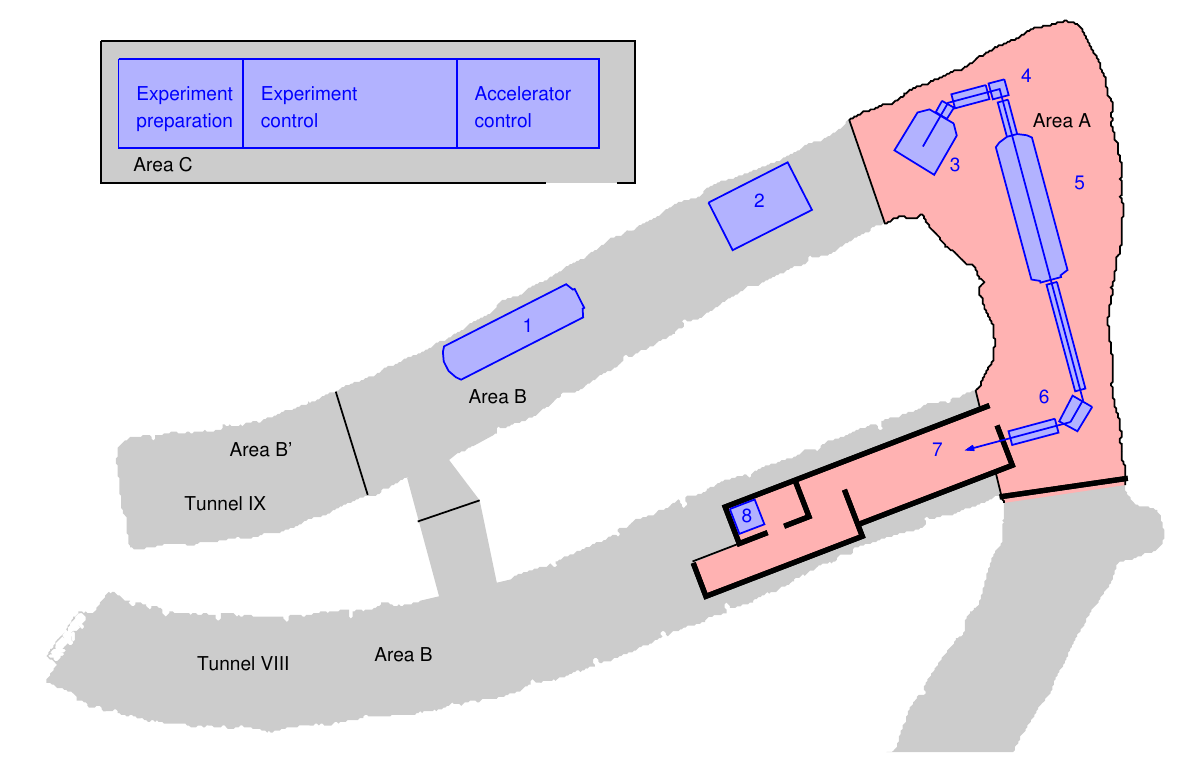}
\caption{\label{fig:Felsenkeller-Plan} Layout of the Felsenkeller tunnels VIII and IX, now under construction: (1) SF$_6$ storage tank. (2) SF$_6$ pumping station. (3) MC-SNICS sputter ion source. (4) Low-energy magnet. (5) 5\,MV Pelletron accelerator with terminal ion source. (6) High-energy magnet. (7) Bunker for in-beam experiments. (8) 150\% HPGe detector, inside bunker for activation experiments. Area C will be located in a surface building a few meters removed from the tunnel mouth.
}
\end{figure*}

\section{Site and background}

Inside the city limits of Dresden, there is a system of nine tunnels shielded from cosmic radiation by 45\,m of hornblende monzonite rock. Since 1982, one tunnel called tunnel IV hosts an underground laboratory for low-radioactivity measurements  \cite{Helbig84-Isotopenpraxis,Koehler09-Apradiso}. This facility has recently been used for a $^{44}$Ti activation study for nuclear astrophysics \cite{Schmidt13-PRC}.

For radiative-capture experiments in nuclear astrophysics, the background in the 4-10\,MeV $\gamma$-energy region is of decisive importance. By moving one and the same detector to several different sites successively, this background has been studied in detail \cite{Szucs12-EPJA,Szucs15-EPJA}. It is found that in an escape-suppressed high-purity germanium (HPGe) detector system, the background in the 6-8\,MeV region is only a factor of 2-4 higher at Felsenkeller than at the deep-underground site LUNA \cite{Szucs12-EPJA}.

Based on measurements of the muon flux, it is estimated that the rock overburden in Felsenkeller tunnels VIII and IX is equivalent to 130\,m of water \cite{Olah16-NPA6}. A complete muon map of the tunnel system is currently under analysis. 

A neutron background measurement is underway, as well. Preliminary data indicate that the overall neutron flux at Felsenkeller is three times higher than at the deep-underground Canfranc laboratory, Spain.

\section{Pelletron accelerator and ultra-low-background HPGe detector}

Two instruments shall be installed in tunnels VIII and IX (Figure~\ref{fig:Felsenkeller-Plan}): First, a 5\,MV Pelletron accelerator that was acquired in 2012 by HZDR and that is currently being readied for installation underground. This accelerator has double charging chains (250\,$\mu$A upcharge current and two ion sources, an external sputter ion source for intensive hydrogen and carbon beams, and an internal radio-frequency ion source to be mounted on the high voltage terminal. 

The second major instrument to be installed in tunnels VIII and IX is a new, large ultra-low background HPGe detector that will be used for offline activation studies and that can also be made available for the purpose of material selection.

\section{Project status}

The project is fully funded, jointly by the two partners HZDR and TU Dresden. Construction work has started in August 2016 with the removal of the old floor from tunnels VIII and IX. The insertion of the Pelletron accelerator tank is planned for April 2017. Construction shall be finished at the end of August 2017. The set-up of the beam lines and commissioning of the accelerator is planned for fall 2017, so that the beam will be available at the end of 2017. 

\section{Access and use}

The new instruments in Felsenkeller shall be made available for two user groups: First, for in-house research by the nuclear astrophysics groups from HZDR and TU Dresden, with current research priorities including helium burning and solar fusion.  Second, a significant amount of beam time will be made available free of charge to outside scientific users, based on the recommendations from an independent advisory body. The detailed access modalities will be announced in due course.

\subsection*{Acknowledgments}

Support by the Helmholtz Association Nuclear Astrophysics Virtual Institute (NAVI, HGF VH-VI-417) and by DFG (TU Dresden Institutional Strategy, "support the best", and INST 269/631-1 FUGG) is gratefully acknowledged.

%\bibliography{Danielsbib}
%\bibliographystyle{jpsj-db}

\end{document}